# A Novel Architecture for Antenna Arrangement in Wireless Cellular CDMA Systems

Hamed Saghaei, *Student Member, IEEE*
Islamic Azad University, Shahrekord branch, Shahrekord, Iran.
Email: saghaei@ieee.org

*Abstract*— **Wise arrangement of antennas is critical in wireless cellular systems for both reduction of co-channel interference (CCI) and increase the quality of service (QoS). In this paper, a novel architecture for antenna arrangement in CDMA wireless cellular systems is presented. In this architecture that we called microzone, every cell is divided into three (or more) zones and information transmission in downlink channel is done by an antenna which is placed at the outer region of the related zone. Also, the transmitting signal by the mobile station (MS) in uplink channel is received by all antennas of the related cell. Analytical calculations of the received signal to noise ratio (SIR) and outage probability for both microzone and used architectures show that proposed architecture has better performance in compared with the used architecture. Also, simulation results confirm lower outage probability in uplink channel for microzone architecture.**

## 1. Introduction

Frequency reuse in wireless cellular systems based on either code division multiple access (CDMA) or frequency/time division multiple access (FDMA-TDMA) results in co-channel interference (CCI) which is one of the major factors that limits the capacity of cellular systems [1]. Co-channel interference arises when the same carrier frequency is used in different cells. In this case, the power spectral density of the desired and interfering signals completely overlap [2]. The generated levels of CCI can be controlled by the use of several techniques such as cell sectoring, smart antennas, power control, discontinuous transmission, effective hand-off algorithms, macroscopic base station (BS) diversity and etc. [3]. The impact of CCI on the radio link can be mitigated using many techniques such as interference cancellation, error control coding, and antenna diversity [4]. Against of commonly used techniques, this paper proposes a novel architecture for the antenna arrangement. This paper is organized as follows

In section 2, cellular architectures are introduced. System model is investigated in section 3. Simulation results are shown in section 4 and section 5 concludes the paper.

## 2. Cellular Architectures

The aim of this section is introducing the used and the microzone architectures. In Fig. 1, the used architecture is shown that it includes many clusters to cover a big area in which each cluster encompasses 1, 3, 5, or 7 cells [5]. Every cell has a base station (BS) which is located at the center of the cell. Cell sectoring may be used which is a way for increasing the system capacity while keeping the cell radius constant. It is a process of replacing one omnidirectional antenna at the BS by several directional antennas. Each of these antennas radiates within a specific sector of the cell. Directional antennas, therefore, minimize interference for a given cell by receiving and transmitting with only a fraction of available co-channel cells, so the QoS will be increased [5]. The degree of interference minimization depends on the amount of used sectors. In general, a sector corresponds to $60°$ or $120°$. Since each sectorized antenna radiates within a specific sector, each sector is allocated a subset of the frequency channels available for the cell. When sectoring is used, the channels in a particular cell are broken down to sectored groups. Fig. 2 shows that $60°$ sectoring reduces the co-channel interference (CCI) by a factor of six, while the $120°$ sectoring reduces it by a factor of three [4]. Fig. 3 shows microzone architecture in which, each of cells is divided into three (or more) zones and each zone has an antenna that is connected by cable or microwave link to the BS of the Cell. Antennas are placed at the outer edge of a cell. Cells are shown by circles, while zones are represented by hexagons circumscribed within each circle. The microzone antennas are designated by black semi-circles and mobile station (MS) is determined by black triangular. When a MS travels within the cell, its transmitted signal in uplink channel is received by all antennas in a cell and by the use of diversity technique, i.e., maximum ratio combining (MRC); so that an acceptable level of the received signal is achieved. Another advantage of this architecture is remaining a MS in the same cell with same frequency band as long as it travels from one zone to another within the cell. Therefore, unlike sectoring, handoff is not required when the MS travel zone to zone in a cell, and only, the BS switches the signal from one zone to another. In this case, the used antennas are directional and the microzone architecture is configured similar to used architecture (i.e., it has 1, 3, 5 or 7 cells per cluster).

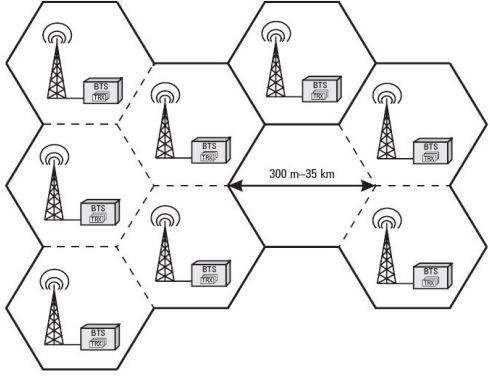

**Fig. 1.** Used architecture [5]

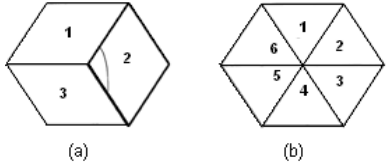

**Fig. 2.** Cell sectoring: (a) 120° sectoring, and (b) 60° sectoring [5].

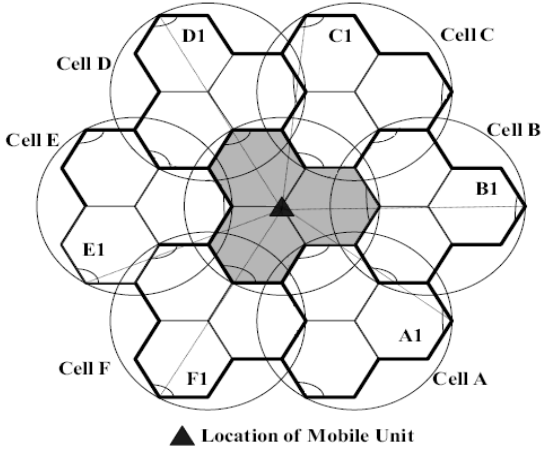

**Fig. 3.** Microzone architecture for one cell per cluster (i.e., frequency reuse factor equals to one).

## 3. System Model

The transmitted signals of all users will be experienced with the channel attenuation that is the most important specification of the radio channel which is shown by $h$ and calculated as follows [6]

$$h = h_p h_s h_m, \quad (1)$$

where $h_p$ represents the path loss that is the large-scale distance-dependent attenuation in the average signal power. $h_s$ shows shadowing loss that is the medium-scale attenuation, which is caused by reflections, refractions and diffractions of the signal from buildings, trees, and rocks. These result in relatively slow variations in the mean signal power and modeled as a log-normally distributed random variable and known as slow fading [6]. $h_m$ represents the multipath fading or fast fading attenuation, which is the rapid fluctuation in the received signal power that is caused by the constructive and destructive addition of the signals that propagate through different paths with different delays from the transmitter to the receiver. In urban environment the number of significant signal paths is typically much larger than rural areas. This is called the *multipath spread*, which is the time between the first occurrence of the transmitted signal at the receiver and the last significant reflection of the same signal at the receiver [7]. If the multipath spread is smaller than the inverse of the bandwidth of the information-bearing signal, i.e., smaller than the duration of a transmitted symbol, then the fading is said to be frequency-non-selective or flat fading [7]. [8] proposes a model for a Rayleigh fading channel with the required spectral properties based on a sum of sinusoids which is used in this paper. Based on these concepts, the channel gain, $g$, is calculated as [5]

$$g = \frac{1}{h} = (A_p d^{-\rho})(10^{\xi/10})(A_f), \quad (2)$$

where $A_p$ depends on the transmission wavelength ($\lambda$), the antenna gain of the transmitter ($g_T$) and the antenna gain of the receiver ($g_r$) that is given by [5]

$$A_p = g_T g_r \lambda^2 / (4\pi)^2, \quad (3)$$

and $d$ is the distance between the transmitter and the receiver, $\rho$ is the path loss exponent with typical values ranging from 2 (free space propagation) to 5 (dense urban areas) [6], $\xi$ is a mean zero Gaussian random variable with the standard deviation between 3 to 8 $dB$ [6], and $A_f$ is assumed to be independent exponentially distributed random variable (In a Rayleigh fading environment, the received signal envelop, $\sqrt{A_f}$ has a Rayleigh distribution and its power, $A_f$, has an exponential distribution [7]). In other words, the received power at the BS from a MS is an exponentially distributed random variable.

Transmitted signals are also polluted by additive white Gaussian noise (AWGN). In CDMA systems, co-channel users' interference power is much greater than the noise power, and then we use interference power instead of users' interference plus noise power in the following analysis. Therefore, the ratio of the transmitted signal power to interference power is called signal to interference ratio (SIR).

### 3. 1. Received SIR in uplink for used architecture based on Fig. 1

If $i$-th MS is served by $b$-th BS, Then the uplink measured SIR for $i$-th MS at time $k$ is calculated as

$$\gamma_{bi}(k) = \frac{g_{bi}(k) p_i(k)}{\left(\sum_{j=1, j \neq i}^{N} g_{bj}(k) p_j(k)\right) + \eta_b} P_G, \quad (4)$$

where $p_i(k)$, $g_{bi}(k)$ and $\eta_b$ are the transmitted power of user $i$, link gain between MS $i$ and BS $b$ at time $k$ (which is an exponentially distributed random variable) and the noise power which is additive white Gaussian noise (AWGN), respectively. $N$ is the number of users that distributed in environment, $P_G$ is the processing gain which equals to $(w/R)$ that $w$ is the total spread spectrum bandwidth occupied by the CDMA signals and $R$ is the bit rate transmitted from MS $i$. Note that in a Rayleigh fading channel, $\gamma_{bi}$ is a random variable with a complex distribution, since it has a ratio of an exponential random variable to a sum of exponential random variables with different means. For reliable connection the received SIR should be more than the target value, i.e., $(\gamma_{bi}(k) \geq \gamma_{bi}^T)$ [2].

We assume that the quality of service (QoS) requested is provided when the SIR exceeds a given threshold ($\gamma^{th}$). The outage probability of the $i$-th user at time $k$ is given by

$O_{used,i}(k) = Prob\{\gamma_{bi}(k) \leq \gamma_i^{th}\}$

$= Prob\left\{\left(\frac{g_{bi}(k)p_i(k)}{\sum_{j=1,j\neq i}^{N} g_{bj}(k)p_j(k) + \eta_b} P_G\right) \leq \gamma_i^{th}\right\}$

$= Prob\left\{g_{bi}(k)p_i(k) \leq \frac{\gamma_i^{th}}{P_G}\left(\sum_{j=1,j\neq i}^{N} g_{bj}(k)p_j(k) + \eta_b\right)\right\}$, (5)

*Lemma1*: Suppose $z_1, z_2, \ldots, z_n$ are independent exponentially distributed random variables with mean ($E(z_i) = 1/Y_i$), and $c$ is a constant, then

$Prob\left(z_1 \leq \sum_{i=2}^{n} z_i + c\right) = 1 - \left[e^{Y_1 c} \prod_{i=2}^{n}\left(1 + \frac{Y_1}{Y_i}\right)\right]^{-1}$ (6)

*Proof*: See Appendix 1.

The mean value of the received power in $b$-th BS from $i$-th MS at time $k$ is given by

$E[g_{bi}(k)p_i(k)] = \bar{g}_{bi}(k)\bar{p}_i(k)$ (7)

Using equation (6), the measured outage probability is calculated as

$O_{used,i}(k) =$
$1 - \left[e^{(\eta_b \gamma_i^{th}/P_G \bar{g}_{bi}(k)\bar{p}_i(k))} \prod_{j\neq i}\left(1 + \frac{\gamma_i^{th} \bar{g}_{bj}(k)\bar{p}_j(k)}{P_G \bar{g}_{bi}(k)\bar{p}_i(k)}\right)\right]^{-1}$ (8)

### 3.2. Received SIR in uplink for microzone architecture based on Fig. 3

In wireless systems, in many situations, there may not be a line-of-sight between the MS and the BS. Therefore, transmitted signals are received from different paths with different delays at the receiver and this causes abrupt increases or decreases in the received signal power. This problem is more troublesome in uplink channel due to the low transmitting power of the MS. In microzone architecture because of using zones in each cell, antenna diversity in the BS is used in uplink channel to overcome this problem. Also, the distance between each pair of antennas is at least several times greater than the transmitted signal wavelength, so that the signal transmission paths would stay independent from each other. Then, if the probability of a received signal from an antenna falls below the threshold, equals $P$, the probability that all the received signals by $L$ antennas simultaneously falls below the threshold, would be $P^L$, which is considerably smaller than $P$. If $i$-th MS is served by $b$-th BS includes an antenna in each zone. Then the uplink measured SIR from 1-th antenna at time $k$ is calculated as

$\gamma_{1i}(k) = \frac{g_{1i}(k)p_i(k)}{\left(\sum_{j=1,j\neq i}^{N} g_{1j}(k)p_j(k)\right) + \eta_1} P_G$, (9)

The received signals by all antennas are combined using MRC algorithm [8] at the BS, in which a specific weight is allocated to each antenna where this weight is given by [8]

$w_{li}(k) = \frac{\sqrt{\gamma_{li}(k)}}{\sum_{l=1}^{L}\sqrt{\gamma_{li}(k)}}$, (10)

where $w_{li}(k)$ is the weighting factor related to the $l$-th antenna at time $k$ and $\gamma_{li}$ is the received SIR of the $l$-th antenna related to zone $l$ and the output of the diversity combiner, $\gamma_{div,i}$, is calculated as

$\gamma_{div,i}(k) = \sum_{l=1}^{L} w_{li}(k)\gamma_{li}(k).$ (11)

Equation (10) shows that using diversity technique, greater value of the received SIR is achievable and possible in microzone architecture.

We assume that the quality of service (QoS) requested is provided when the output of the diversity combiner, $\gamma_{div,i}$ exceeds a given threshold ($\gamma^{th}$). The outage probability of the $i$-th user at time $k$ is given by

$O_{Micro,i}(k) = Prob\{\gamma_{div,i}(k) \leq \gamma_i^{th}\}$

$= Prob\left\{\sum_{l=1}^{L}\left(\frac{w_{li}(k)g_{li}(k)p_i(k)}{\sum_{j=1,j\neq i}^{N} g_{lj}(k)p_j(k) + \eta_l} P_G\right) \leq \gamma_i^{th}\right\}$ (12)

Microzone architecture has some advantages that are listed as follows

1) The received SIR is increased by the use of diversity technique.
2) If users uniformly distributed in a cell then a large number of them are distributed at the regions near to antennas of a cell in compared with the used architecture. Therefore, by the use of microzone architecture, the received SIR is also increased.
3) An increase in the received SIR will decrease the outage probability and finally, network capacity and performance will be increased.
4) Handoff is not required when the MS travel in a cell.

Because of using diversity technique, antenna arrangement, and above mentioned reasons, $O_{Micro,i}(k) \ll O_{used,i}(k)$ that is shown and confirmed by the simulation results in the next section.

## 4. Simulation Results

In this section, the simulations are done for used and microzone architectures to investigate the system performance. Based on Fig. 2.a, for used architecture, the $120°$ cell sectoring is assumed. Also, the gain of each sectorized antenna is supposed to be $0\ dB$, which is kept constant during the simulations. In the microzone system, it is not possible to use omnidirectional antennas, because the antennas must be located at the outer edge of any cell and cover only the related zones. However, both $60°$ and $120°$ antennas are employed in the microzone system but in these simulations, we use only $120°$ antennas to have the same conditions with the used architecture. The BS antennas for the uplink are assumed to be identical. The cluster sized is assumed by the value of 1. The number of users that are uniformly distributed over a cell is 40. Each user generates its data at a constant rate of $45\ kb/s$. The generated data is modulated using BPSK and then, using corresponding spreading sequence, the data is spread with a rate of $3.8\ Mchip/s$ (therefore, the processing gain will be $19.3\ dB$). The spreading signals are transmitted in the uplink channel. Here, the SIR threshold is supposed to be about $0\ dB$ for every user. The transmitting signals are corrupted by AWGN with a mean of zero and a standard deviation of $5\ dB$ [9]. Here we ignore the effect of Doppler frequency. The path loss exponent $\rho$ equals to 4.

Fig. 4 shows the outage probability versus the SIR for used and microzone architectures and confirms the better performance of microzone architecture.

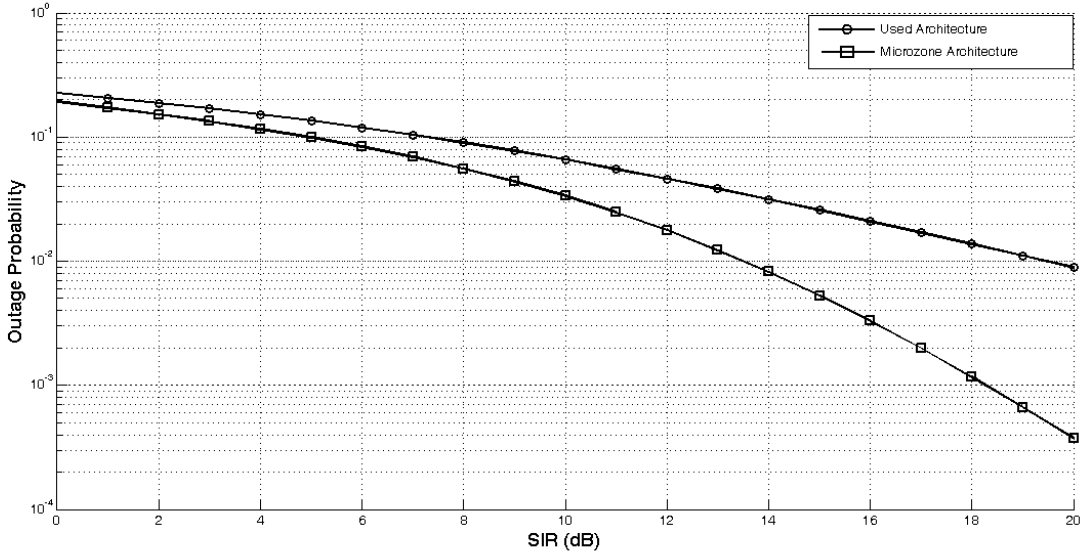

Fig. 4. Outage probability versus SIR for both microzone and used architectures

## 5. Conclusion

In this paper, the used architecture was introduced and a novel architecture for antenna arrangement that called microzone was proposed. Also, the mathematical analyses on outage probability for both of architecture were calculated and shown better performance of the proposed architecture. Moreover, the simulation results confirmed the lower outage probability of the microzone in compared with the used architecture which results in higher QoS, and lower bit error rate at the BS.

## Appendix 1:

In this section, we prove the equation (6). Suppose $z_1, z_2, \ldots, z_n$ are independent exponentially distributed random variables with mean $(E(z_i) = 1/\Upsilon_i)$, and $c$ is a positive constant, then

$$Prob\left(z_1 \leq \sum_{i=2}^{n} z_i + c\right) = 1 - \left[e^{\Upsilon_1 c} \prod_{i=2}^{n}\left(1 + \frac{\Upsilon_1}{\Upsilon_i}\right)\right]^{-1},$$

To prove this, we note that

$$Prob\left(z_1 \leq \sum_{i=2}^{n} z_i + c\right)$$

$$= \int_0^\infty \ldots \int_0^\infty \int_0^{\sum_{i=2}^n z_i + c} f(z_1)f(z_2)\ldots f(z_n)\, dz_1 dz_2 \ldots dz_n$$

$$= \int_0^\infty \ldots \left(\int_0^\infty \left(\int_0^{\sum_{i=2}^n z_i + c} f(z_1)\, dz_1\right) f(z_2) dz_2\right)\ldots f(z_n) dz_n$$

$$= \int_0^\infty \ldots \left(\int_0^\infty \left(\int_0^{\sum_{i=2}^n z_i + c} \Upsilon_1 e^{-\Upsilon_1 z_1} dz_1\right) f(z_2) dz_2\right)\ldots f(z_n) dz_n$$

$$= \int_0^\infty \ldots \left(\int_0^\infty \left(1 - e^{-\Upsilon_1(\sum_{i=2}^n z_i + c)}\right) \Upsilon_2 e^{-\Upsilon_2 z_2} dz_2\right)\ldots f(z_n) dz_n$$

$$= \int_0^\infty \ldots \left(\int_0^\infty \Upsilon_2 \bigl(e^{-\Upsilon_2 z_2} - e^{-\Upsilon_1(\sum_{i=3}^n z_i + c) - (\Upsilon_1 + \Upsilon_2) z_2}\bigr) dz_2\right)\ldots f(z_n) dz_n$$

$$= \int_0^\infty \ldots \left(\int_0^\infty \left(1 - \frac{\Upsilon_2 e^{-\Upsilon_1(\sum_{i=3}^n z_i + c)}}{\Upsilon_1 + \Upsilon_2}\right) f(z_3)\, dz_3\right)\ldots f(z_n) dz_n$$

$$= \int_0^\infty \ldots \left(\int_0^\infty \left(1 - \frac{\Upsilon_2 \Upsilon_3 e^{-\Upsilon_1(\sum_{i=4}^n z_i + c)}}{(\Upsilon_1 + \Upsilon_2)(\Upsilon_1 + \Upsilon_3)}\right) f(z_4)\, dz_4\right)\ldots f(z_n) dz_n$$

$$= \int_0^\infty \ldots \left(\int_0^\infty \left(1 - \frac{\Upsilon_2 \Upsilon_3 e^{-\Upsilon_1(\sum_{i=4}^n z_i + c)}}{(\Upsilon_1 + \Upsilon_2)(\Upsilon_1 + \Upsilon_3)}\right) f(z_4)\, dz_4\right)\ldots f(z_n) dz_n$$

$$= \left(1 - e^{-\Upsilon_1 c} \prod_{i=2}^{n}\left(\frac{\Upsilon_i}{\Upsilon_1 + \Upsilon_i}\right)\right)$$

$$= 1 - \left[e^{\Upsilon_1 c} \prod_{i=2}^{n}\left(1 + \frac{\Upsilon_1}{\Upsilon_i}\right)\right]^{-1}, \qquad (12)$$